\documentclass[a4paper]{article}
\usepackage{epsfig}
\begin{document}
\begin{titlepage}
\title{Cryptographic Pseudo-Random Sequences from the Chaotic H\'{e}non Map}
\author{Madhekar Suneel\\ \\
				Department of Electrical Communication Engineering\\
				Indian Institute of Science, Bangalore - 560 012.\\
				(address for correspondence)\\
				\& \\
				PGAD, Defence Research and Development Organization\\
				(Ministry of Defence, Government of India)\\
				DRDL Complex, Kanchanbagh, Hyderabad - 500 058.\\
				(permanent address)\\ \\
				suneel@ieee.org}
\date{}
\maketitle              

\end{titlepage}
\begin{abstract}
\noindent A scheme for pseudo-random binary sequence generation based on the two-dimensional discrete-time H\'{e}non map is proposed.  Properties of the proposed sequences pertaining to linear complexity, linear complexity profile, correlation and auto-correlation are investigated.  All these properties of the sequences suggest a strong resemblance to random sequences.  Results of statistical testing of the sequences are found encouraging.  An attempt is made to estimate the keyspace size if the proposed scheme is used for cryptographic applications.  The keyspace size is found to be large and is dependent on the precision of the computing platform used.
\end{abstract}
\section{Introduction}
Pseudo-random number sequences are useful in many applications including monte-carlo simulation, spread spectrum communications, steganography and cryptography.  Conventionally, pseudo-random sequence generators based on linear congruential methods and feedback shift-registers are popular \cite{knuth98}.  For cryptographic applications, several algorithms such as ANSI X9.17 and FIPS 186 are found to be popular \cite{menezes97}.  In recent times, several researchers have been exploring the idea of using chaotic dynamical systems for this purpose \cite{falcioni06,kocarev01,woodcock98}.  The random-like, unpredictable dynamics of chaotic systems, their inherent determinism and simplicity of realization suggests their potential for exploitation as pseudo-random number generators.

Cryptographic schemes based on chaos have been classified as~~1) \emph{discrete-time discrete-value} schemes~~2) \emph{discrete-time continuous-value} schemes and~~3) \emph{continuous-time continuous-value} schemes \cite{dachselt01}.  All conventional cryptographic schemes act on discrete symbols such as bits, integers, characters and symbols and can be grouped into the first category, that of discrete-time discrete-value schemes.  In this paper, a scheme for obtaining a pseudo-random binary sequence from the two-dimensional chaotic H\'enon map is explored.  Needless to say, since the sequence is a discrete-time discrete-value signal, any cryptographic application of the same can be grouped under the first of the three classes mentioned above.

\section{The H\'enon Map}
\begin{figure}[ht]
\begin{center}
\includegraphics[scale = 0.4]{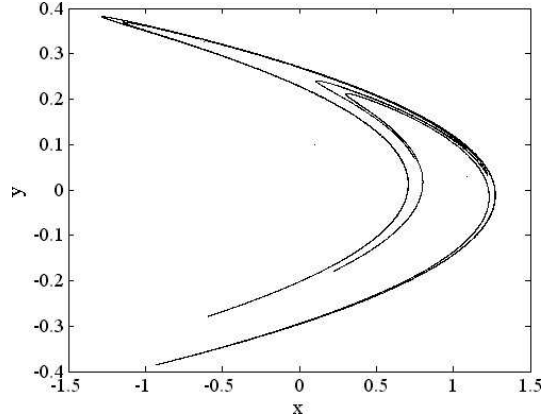}
\caption{The Strange Attractor of H\'{e}non}
\label{fighenonstate}
\end{center}
\end{figure}

An $N$-dimensional discrete-time dynamical system is an iterative map $f : \Re^{N} \rightarrow \Re^{N}$ of the form
\begin{equation}
X_{k+1} = f\left(X_{k}\right) \enspace ,
\label{eqdynasystem}
\end{equation}
where $k = 0, 1, \ldots$ is the discrete time and $X \in \Re^{N}$ is the state.  Starting from $X_0$, the initial state, repeated iteration of (\ref{eqdynasystem}) gives rise to a series of states known as an orbit.  An example is the H\'{e}non map, a two-dimensional discrete-time nonlinear dynamical system represented by the state equations \cite{henon76,alligood97,peitgen04,kumar96,crutchfield86,gleick97}
\begin{equation}
\begin{array}{rcl}
x_{k+1} &=& -\alpha x_k^2 + y_k +1 \enspace , \\
y_{k+1} &=& \beta x_k \enspace.
\end{array}
\label{eqhenonmap}
\end{equation}
Here, $\left(x,y\right)$ is the two-dimensional state of the system.  The state-plane diagram for $\alpha = 1.4$ and $\beta = 0.3$ for this map is shown in Fig. \ref{fighenonstate}.  The diagram is a strange attractor popularly known as the H\'{e}non attractor.

In this paper, the H\'{e}non map shall be considered a representative example of 2-dimensional chaotic maps for the generation of pseudorandom sequences.

\section{Pseudorandom Sequence Generation Scheme}
Generating a pseudorandom binary sequence from the orbit of a chaotic map essentially requires mapping the state of the system to $\left\{0, 1\right\}$.  For the H\'{e}non map, consider the two bits $b_x$ and $b_y$ derived respectively from the $x$ and $y$ state-variables as follows:
\begin{eqnarray}
  b_x &=& \left \{ 
  			\begin{array}{rll}
  			1 \quad & \mbox{if} & x > \tau_x \enspace ; \\
				0 \quad & \mbox{if} & x \le \tau_x \enspace .
				\end{array}
				\right.			\\
	b_y &=& \left \{ 
  			\begin{array}{rll}
  			1 \quad & \mbox{if} & y > \tau_y \enspace ; \\
				0 \quad & \mbox{if} & y \le \tau_y \enspace .
				\end{array}
				\right.	
\end{eqnarray}
Here, $ \tau_x $ and $ \tau_y $ are appropriately chosen threshold values for state-variables $x$ and $y$.  $\tau_x$ should be chosen such that the likelihood of $x > \tau_x$ is equal to that of $x \le \tau_x$.  The median of a large set of numbers has precisely this property. Therefore, we choose $\tau_x$ as the median of a large number $\left(T\right)$ of consecutive values of $x$.  Similarly, we assign to $\tau_y$, the value of the median of $T$ consecutive values of $y$.  Thus, two streams of bits $S_x = \left\{b_x^i\right\}_{i = 1}^{\infty}$ and $S_y = \left\{b_y^i\right\}_{i = 1}^{\infty}$ are obtained from the map.  Consider the bit-stream $B_x$ formed by choosing every $P$th bit of $S_x$, i.e., $B_x = \left\{b_x^{Pi}\right\}_{i = 1}^{\infty}$.  Consider the similarly formed bit-stream $B_y = \left\{b_y^{Pi}\right\}_{i = 1}^{\infty}$.  Let us denote the $j$th bit of these two sequences respectively as $B_x\left(j\right)$ and $B_y\left(j\right)$.  Then, the pseudo-random output bit $O$ is chosen as per the following rule:
\begin{equation}
O\left(j\right) = \left\{
												\begin{array}{rll}
												B_x\left(j\right) \quad & \mbox{if} & B_y\left(j-2\right) = 0 \mbox{ and } B_y\left(j-1\right) = 0 \enspace ; \\
												\overline{B}_x\left(j\right) \quad & \mbox{if} & B_y\left(j-2\right) = 0 \mbox{ and } B_y\left(j-1\right) = 1 \enspace ; \\
												B_y\left(j\right) \quad & \mbox{if} & B_y\left(j-2\right) = 1 \mbox{ and } B_y\left(j-1\right) = 0 \enspace ; \\
												\overline{B}_y\left(j\right) \quad & \mbox{if} & B_y\left(j-2\right) = 1 \mbox{ and } B_y\left(j-1\right) = 1 \enspace .
												\end{array}
									\right.
\end{equation}
Here, $\overline{B}_x$ and $\overline{B}_y$ respectively denote the logical inverse of $B_x$ and $B_y$.  For $j = 0$, $B_y\left(-2\right)$ and $B_y\left(-1\right)$ can arbitrarily be assumed to be $0$.

The author has found that generated pseudorandom sequences have good statistical properties when $P$ is large.  The author has used $P$ between 75 and 5000 depending on the available time for computation and the length of the sequence required.  In this paper, sequences generated using this method are called \emph{H\'enon map sequences}.

\section{Linear Complexity Properties}
A Linear Feedback Shift Register (LFSR) is said to \emph{generate} an $N$-bit sequence $W$ if for some initial state, the first $N$ bits of the output sequence of the LFSR are the same as $W$ \cite{golomb64,golomb82}.  The length of the shortest LFSR that generates $W$ is known as its \emph{Linear Complexity}.

The author has measured the linear complexity of a large number of even-length H\'{e}non map sequences using the Berlekamp-Massey Algorithm (BMA) \cite{massey69}.  The linear complexities obtained for each sequence-length were found to follow a certain probabilistic pattern.  In particular, the probability of the linear complexity $C$ of an $N$-bit sequence being equal to $c \left(c < N\right)$, when $N$ is even, was found to be very close to
\begin{equation}
P\left(C = c\right) = \left\{ \begin{array}{lll}
															\left(0.5\right)^{N-2c+1} & \quad \mbox{if } & c \le N/2 \enspace ; \\
															\left(0.5\right)^{2c-N} & \quad \mbox{if } & c > N/2 \enspace .
															\end{array}
											\right.	
\label{eqconjeven}																						
\end{equation}

\begin{figure}[ht]
\begin{tabular*}{\textwidth}{@{\extracolsep{\fill}}cc}
\includegraphics[scale = 0.4]{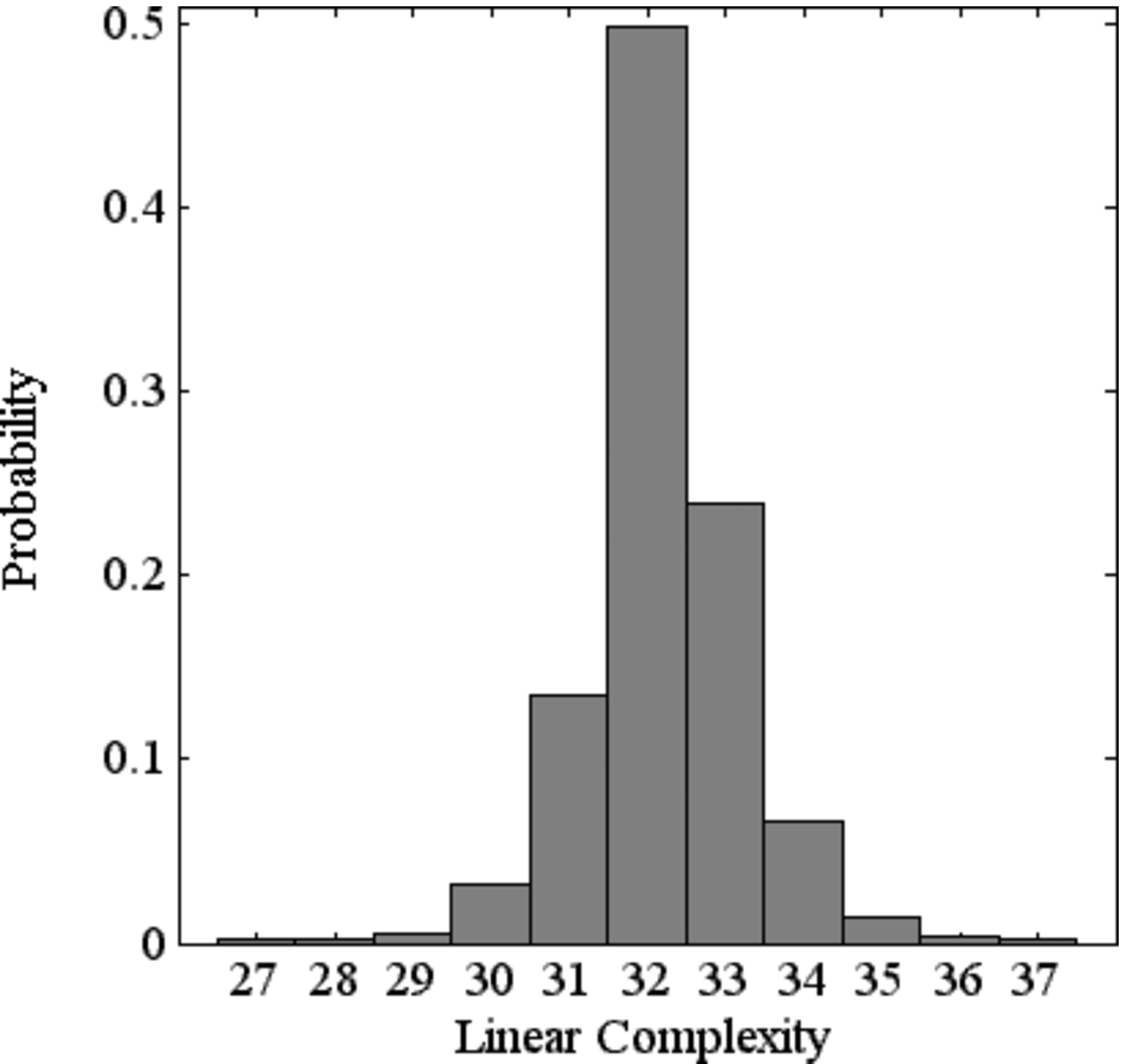}
&
\includegraphics[scale = 0.4]{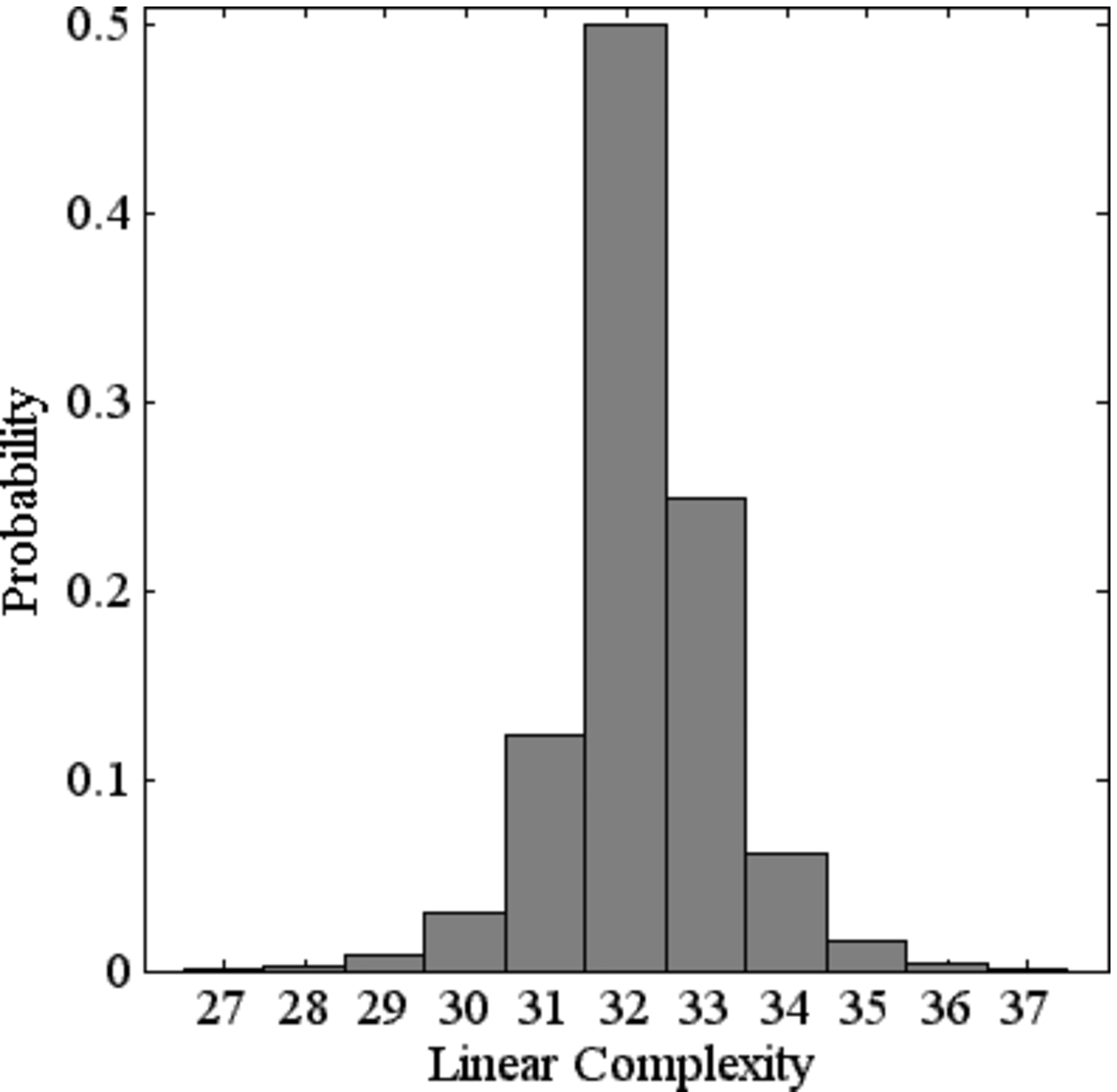}
\\
a)~~experimental & b)~~as per (\ref{eqconjeven})
\end{tabular*}
\caption{Probability Distribution of Linear Complexity: 64-bit sequences}
\label{figconjeven}
\end{figure}

To illustrate the correctness of this conjecture, the experimentally determined distribution of linear complexities for 64-bit H\'{e}non map sequences is shown in figure \ref{figconjeven} a) alongside the conjectured distribution in figure \ref{figconjeven} b) that has been computed using (\ref{eqconjeven}).  The mean value of linear complexities obtained in this experiment was found to be 32.2083.  The expectation of linear complexity for a 64-bit random sequence is 32.2222 and is found to be very close to that of the H\'enon map sequences \cite{menezes97}.  The variance of linear complexities of 64-bit sequences obtained by experiment was 1.0811 against 1.0617 for random sequences.

Similarly, the probability of the linear complexity $C$ assuming the value $c \left(c < N\right)$ when $N$ is odd was found to be very close to
\begin{equation}
P\left(C = c\right) = \left\{ \begin{array}{lll}
															\left(0.5\right)^{N-2c+1} & \quad \mbox{if } & c < \left(N+1\right)/2 \enspace ; \\
															\left(0.5\right)^{2c-N} & \quad \mbox{if } & c \ge \left(N+1\right)/2 \enspace .
															\end{array}
											\right.	
\label{eqconjodd}																						
\end{equation}

\begin{figure}[ht]
\begin{tabular*}{\textwidth}{@{\extracolsep{\fill}}cc}
\includegraphics[scale = 0.4]{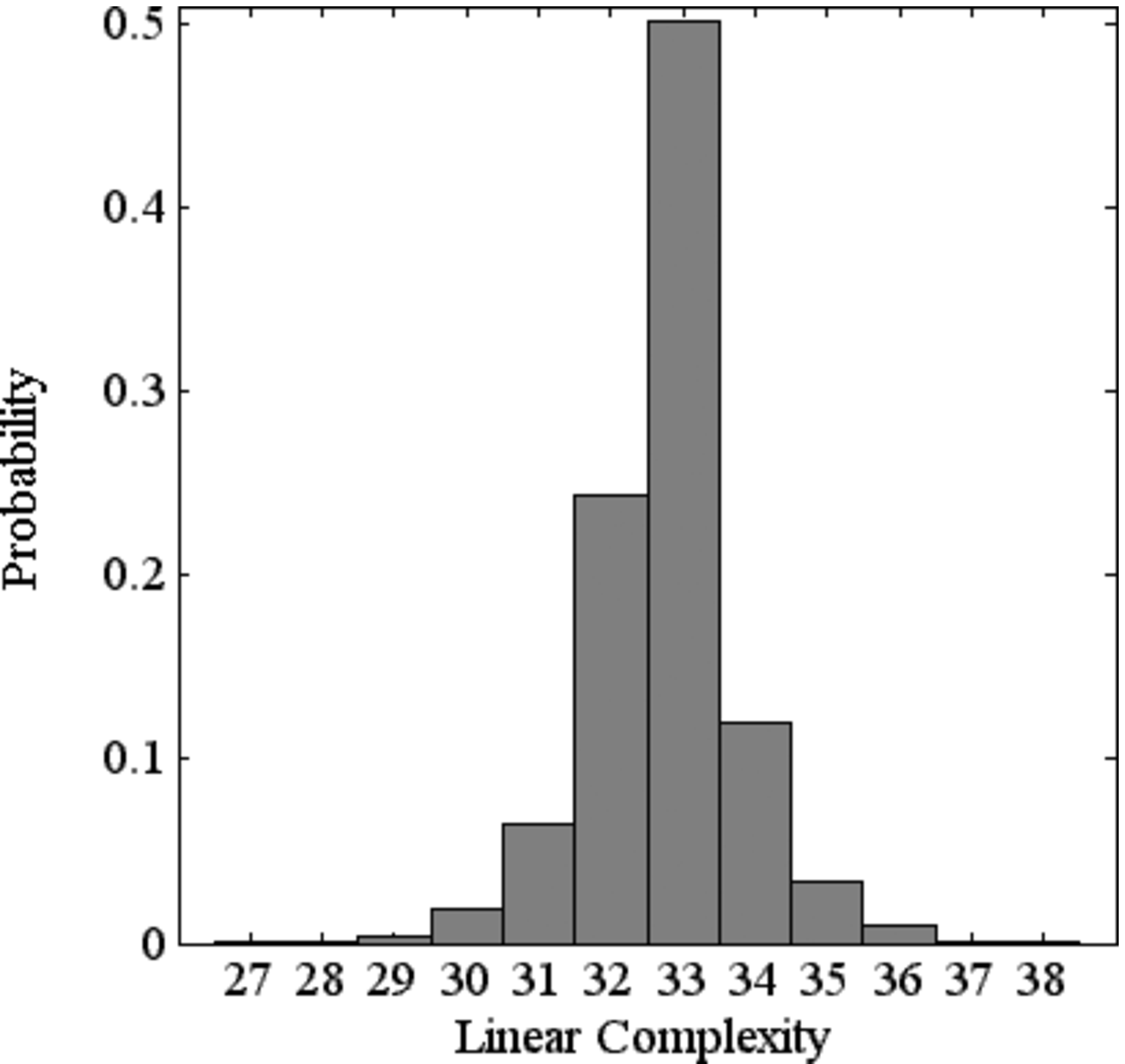}
&
\includegraphics[scale = 0.4]{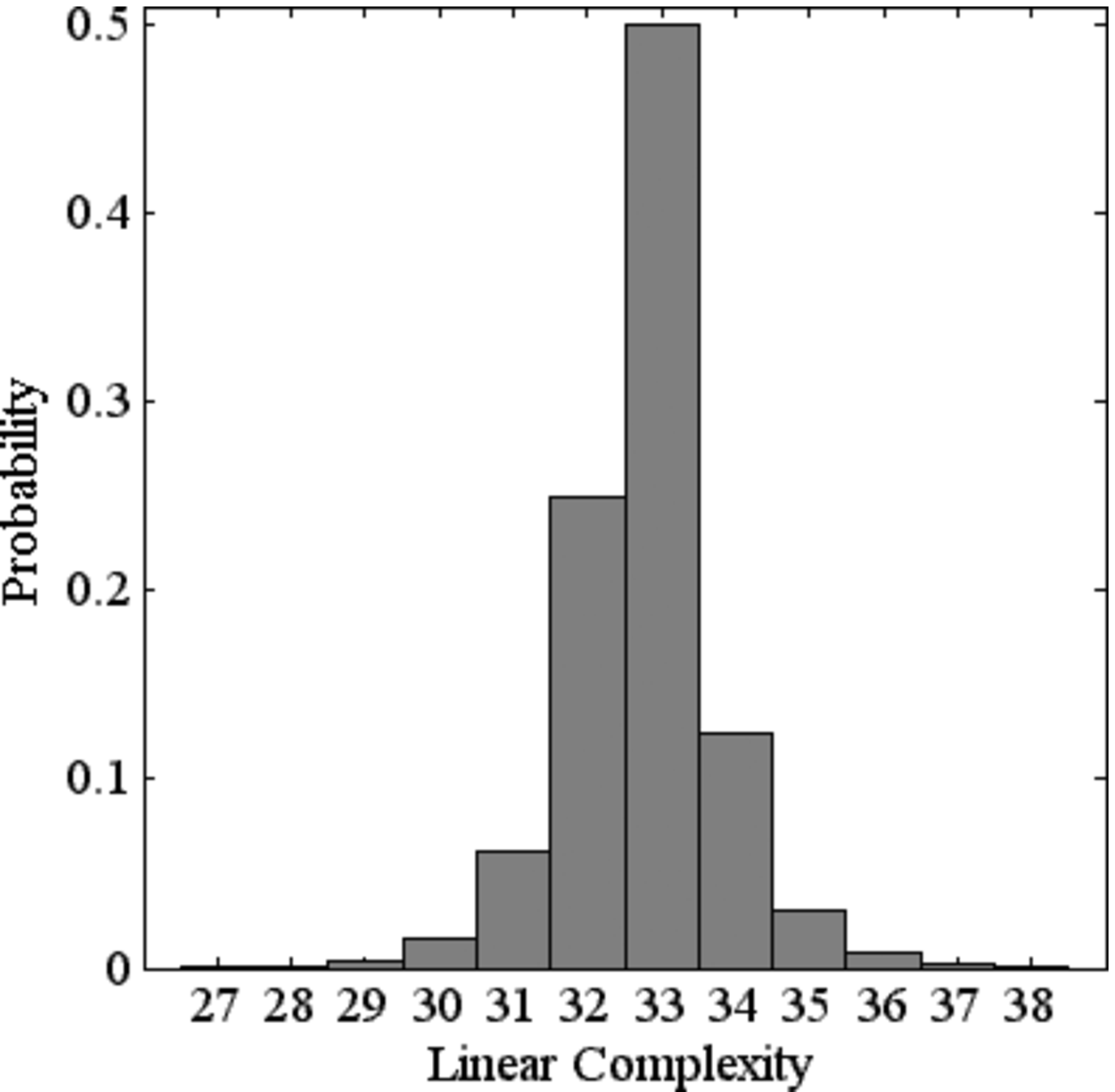}
\\
a)~~experimental & b)~~as per (\ref{eqconjodd})
\end{tabular*}
\caption{Probability Distribution of Linear Complexity: 65-bit sequences}
\label{figconjodd}
\end{figure}

Again, in fig. \ref{figconjodd} a), the experimental distribution is shown along with the experimentally determined distribution in fig. \ref{figconjodd} b) for 65-bit H\'enon map sequences.  The experimentally measured mean of linear complexities of these sequences was 32.7663 against the expected 32.7778 for random sequences.  Also, the measured variance of linear complexities stands at 1.1177 against 1.0617 for random sequences \cite{menezes97}.

\begin{figure}[ht]
\begin{center}
\includegraphics[scale = 0.4]{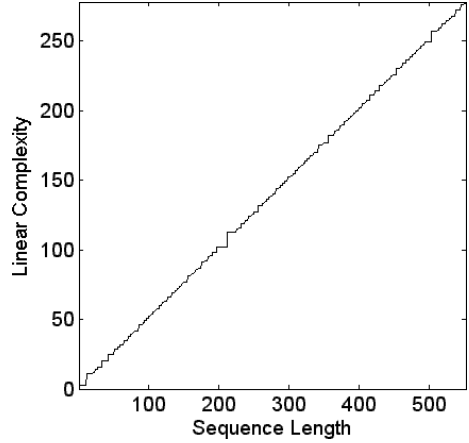}
\caption{Linear Complexity Profile}
\label{figlcprofile}
\end{center}
\end{figure}

Let $W$ be an $N$-bit binary sequence.  Let $W_i \left(i=1,2,\ldots,N\right)$ denote the subsequence of $W$ consisting of its first $i$ bits. Let $C_i$ denote the linear complexity of $W_i$.  Then the sequence of linear complexities $\left(C_1, C_2, C_3, \ldots, C_N\right)$ is known as the linear complexity \emph{profile} of $W$.  For random sequences, the linear complexity profile is expected to be very close to the $C_i=i/2$ line (Menezes et al 1997).  The linear complexity profile of a sample 553-bit sequence was determined using the Berlekamp-Massey Algorithm.  The obtained profile is shown in fig. \ref{figlcprofile} and can be seen to be very close to the $C_i=i/2$ line.

\section{Correlation Properties}
Consider two $N$-bit binary sequences $U$ and $V$.  Let $A$ be the number of bit-by-bit agreements between the two.  The number of bit-by-bit disagreements must be $D = \left(N-A\right)$.  Then the correlation $\theta$ of the two sequences is defined as
\begin{equation}
\theta \left(U,V\right) = \left(A - D\right)/N \enspace .
\label{eqcorrelation}
\end{equation}
\newtheorem{theorem}{Theorem}
\begin{theorem}
If $\;U$ and $V$ are two $N$-bit random binary sequences, the probability of their correlation $\theta$ assuming the value $\Theta$ is given by
\begin{equation}
\label{eqnrandcorr}
P\left(\theta = \Theta\right) = \sqrt{\frac{2}{N \pi}}\;e^{-N \Theta^2 / 2}
\end{equation}
when
\[
\Theta \in \left\{ \begin{array}{ll}
										\left\{ 0,\pm{\frac{2}{N}},\pm{\frac{4}{N}},\ldots,\pm{1} \right\}& \mbox{for even } N, \\
										\left\{ \pm{\frac{1}{N}}, \pm{\frac{3}{N}}, \pm{\frac{5}{N}},\ldots,\pm{1} \right\}& \mbox{for odd } N.
										\end{array}
						\right.
\]
The probability is zero for other values of $\Theta$.
\end{theorem}

\noindent \textit{\textbf{Proof:}}  Since the probability of occurrence of a one is the same as the probability of occurrence of a zero in a truly random sequence, the probability of occurrence of an agreement $\left(p = 0.5\right)$ is the same as the probability of occurrence of a disagreement $\left(1-p = 0.5\right)$ in a pair of such sequences.  Therefore, the number of agreements $A$ in a pair of such sequences is a random variable that follows the binomial distribution with mean $\mu = N/2$ and standard deviation $\sigma = \sqrt{N}/2$.  Therefore
\begin{equation}
P\left(A = r\right) = \frac{^N\!C_r}{2^N} \enspace .
\label{eqnrandprob}
\end{equation}
Applying the Normal approximation to the binomial distribution \cite{keeping62,ramasubramanian97},
\begin{equation}
\label{eqnnormapproxprob}
P\left(A = r\right) \approx \sqrt{\frac{2}{N \pi}}\;e^{-2\left(r-N/2\right)^2/N} \enspace .
\end{equation}
The correlation $\theta$ is related to the number of agreements $A$ as
\begin{equation}
\label{eqnthetaa}
\theta = \frac{2A}{N} - 1.
\end{equation}
As $A \in \left\{0,1,2,\ldots,N \right\}$, $\theta \in \left\{-1, \left(\frac{2}{N}-1\right), \left(\frac{4}{N}-1\right), \ldots,1 \right\}$.  Therefore, $\theta \in \left\{ 0,\pm{\frac{2}{N}},\pm{\frac{4}{N}},\ldots,\pm{1} \right\}$ for even $N$ and $\theta \in \left\{ \pm{\frac{1}{N}}, \pm{\frac{3}{N}}, \pm{\frac{5}{N}},\ldots,\pm{1} \right\}$ for odd $N$.  Clearly, the probability of $\theta$ assuming any other value is zero.
By (\ref{eqnnormapproxprob}) and (\ref{eqnthetaa}), if $\Theta$ belongs to the above set of valid values,
\begin{equation}
P\left(\theta = \Theta\right) = \sqrt{\frac{2}{N \pi}}\;e^{-N \Theta^2 / 2}
\end{equation}
which completes the proof.

\begin{figure}[ht]
\begin{center}
\includegraphics[scale = 0.4]{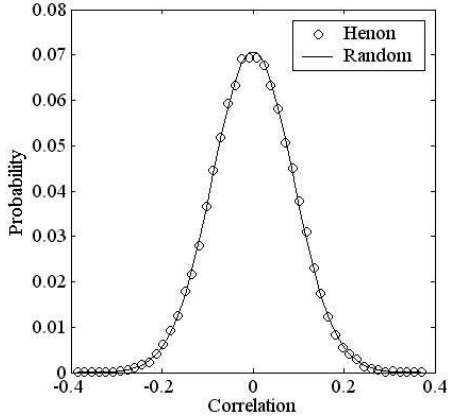}
\caption{Probability Distribution of Correlation.  The legend \emph{Random} is used for a plot of (\ref{eqnrandcorr})}
\label{figcorrelation}
\end{center}
\end{figure}

The correlation between pairs of H\'enon map sequences was experimentally determined and the probability distribution was found to be very close to that of (\ref{eqnrandcorr}).  As an illustration, the probability distribution for 127-bit H\'enon map sequences and the expected distribution for random sequences are shown in Fig. \ref{figcorrelation}.

\begin{figure}[ht]
\begin{center}
\includegraphics[scale = 0.4]{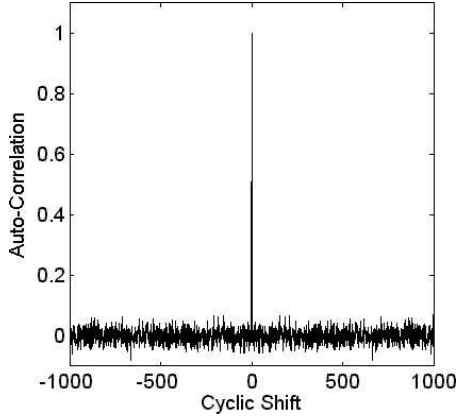}
\caption{Auto-Correlation Function}
\label{figautocorr}
\end{center}
\end{figure}

The correlation of a sequence with a cyclic-shift of itself is known as its \emph{cyclic auto-correlation}.  Let $W$ be an $N$-bit sequence and $W^{j}$ denote $W$ cyclically right-shifted by $j$ bits.  Then the cyclic auto-correlation \emph{function} of $W$ is defined as $R\left(j\right) = \theta\left(W,W^{j}\right)$.  The (cyclic) auto-correlation function of a random sequence is expected to be unity at $j=0$ and close to zero at all other values of $j$.  This indeed was found to be the case with H\'enon map sequnces.  The auto-correlation function of a 2000-bit H\'enon map sequence is shown in fig. \ref{figautocorr}.  In the figure, a negative value of shift signifies cyclic left-shifting by an amount equal to the magnitude.

\section{Statistical Testing}
A number of sequences were generated by the algorithm and were subjected to statistical tests. The tests carried out were Menezes et al's basic tests of randomness \cite{menezes97}, FIPS 140-1 \cite{FIPS-140-1} recommended battery of tests and National Institute of Standards and Technology (NIST) battery of tests \cite{rukhin01}.

Menezes et al have proposed a set of five basic tests consisting of frequency test, serial test, poker test, runs test and auto-correlation test.  While the first four tests were carried out once for a given sequence, the auto-correlation test was carried out for all possible shifted-versions of the sequence.  The result of each test is a test statistic which is compared with a threshold value (one-sided test).  The test results in a failure if the threshold is exceeded. The tests were carried out at a significance level of 0.01, hence 1\% of failures were expected even for random sequences.

FIPS 140-1 recommends a set of statistical tests for cryptographic random/ pseudorandom number generators.  The tests are carried out on a bit-stream of 20,000 bits.  The recommended tests are monobit, poker, runs and long-run tests.  Each test is a two-sided test where a test-statistic is required to lie within an interval.  Though FIPS 140-1 has subsequently been superceded by FIPS 140-2 \cite{FIPS-140-2}, since the latter standard does not recommend any statistical tests for cryptographic random/ pseudorandom number generators, the author has used the older recommendations of FIPS 140-1 to evaluate H\'enon map sequences.

NIST has forumulated a statistical test suite for random/ pseudo-random number generators to be used for cryptographic applications.  The suite consists of a battery of sixteen tests namely 1) Frequency (monobit) test, 2) Frequency test within a block, 3) Runs test, 4) Test for longest run of ones in a block, 5) Binary matrix rank test, 6) Discrete Fourier Transform (spectral) test, 7) Non-overlapping template matching test, 8) Overlapping template matching test, 9) Maurer's universal statistical test, 10) Lempel-Ziv compression test, 11) Linear complexity test, 12) Serial test, 13) Approximate entropy test, 14) Cumulative sums test, 15) Random excursions test and 16) Random excursions variant test.

\section{Statistical Test Results}
\subsection{Menezes et al's basic tests of randomness}
Two 128-bit sequences $R_1$ and $R_2$ were generated using parameters shown in table \ref{tab:MenezesParameters}.  The notation used for test-statistics is the same as those in (Menezes et al 1997). The statistic $X_5$ of the auto-correlation test is a function of the value of the relative circular shift.  In this case, the value of the relative circular shift is shown in brackets.  For example, the statistic obtained by running the auto-correlation test on the sequence and its 3-shifted version is denoted by $X_5\left(3\right)$.  A similar notation is used for the statistic $X_3$ of the poker test where $X_3$ depends on the length of the non-overlapping sub-sequences.  The results of the tests on the sequences are shown in table \ref{tab:Menezes}.
\begin{table}[ht]
		\begin{center} 
		\begin{tabular}{|l|l|l|}
		\hline
			\textbf{Parameter}	&					$\mathbf{R_1}$ & $\mathbf{R_2}$ \\ \hline
			$\alpha$ &	1.40 &	1.20 \\
			$\beta$ &		0.30 &	0.30 \\
			$x_0$ &			-0.75 &	-0.75 \\
			$y_0$ &			-0.02 &		0.32 \\
			$B_y\left(-2\right)$ & 0 & 0 \\
			$B_y\left(-1\right)$ & 1 & 1 \\
			$P$ & 24 & 24 \\
			$T$ & 1000 & 1000 \\ \hline
		\end{tabular}
	\end{center}
	\caption{Parameters used for Menezes' test samples}
	\label{tab:MenezesParameters}
\end{table}
\begin{table}[p]
		\begin{center} \begin{scriptsize}
		\begin{tabular}{|l|l|l|l|l|l|}
		\hline
			\textbf{Statistic}	&	 \textbf{Expected}		&	$\mathbf{R_1}$ & \textbf{Result $\mathbf{\left(R_1\right)}$} & $\mathbf{R_2}$ & \textbf{Result $\mathbf{\left(R_2\right)}$} \\ \hline
			$X_1$ & $< 6.634897$ &	0.125000 &	Pass & 6.125000 & Pass \\
			$X_2$ &	$< 9.210340$ &	0.213583 &	Pass & 6.245079 & Pass \\
			$X_3\left(2\right)$ &	$< 11.344867$ &		4.625000 & Pass &	7.625000 & Pass \\
			$X_3\left(3\right)$ &	$< 18.475307$ &		1.809524 & Pass &		9.428571 & Pass \\
			$X_4$ & $< 9.210340$ & 0.890713 & Pass & 5.373956 & Pass \\
			$X_5\left(1\right)$ & $\left | X_5\left(\cdot\right) \right | < 2.326348$ & 0.443678 & Pass & -0.088736 & Pass\\
$X_5\left(2\right)$ & $\left | X_5\left(\cdot\right) \right | < 2.326348$ & 0.178174 & Pass & 1.069045 & Pass\\
$X_5\left(3\right)$ & $\left | X_5\left(\cdot\right) \right | < 2.326348$ & 0.268328 & Pass & -0.98387 & Pass\\
$X_5\left(4\right)$ & $\left | X_5\left(\cdot\right) \right | < 2.326348$ & -1.257237 & Pass & -0.179605 & Pass\\
$X_5\left(5\right)$ & $\left | X_5\left(\cdot\right) \right | < 2.326348$ & -0.631169 & Pass & -0.811503 & Pass\\
$X_5\left(6\right)$ & $\left | X_5\left(\cdot\right) \right | < 2.326348$ & 0 & Pass & -0.724286 & Pass\\
$X_5\left(7\right)$ & $\left | X_5\left(\cdot\right) \right | < 2.326348$ & 0.272727 & Pass & 0.272727 & Pass\\
$X_5\left(8\right)$ & $\left | X_5\left(\cdot\right) \right | < 2.326348$ & -0.730297 & Pass & -0.912871 & Pass\\
$X_5\left(9\right)$ & $\left | X_5\left(\cdot\right) \right | < 2.326348$ & -0.09167 & Pass & 0.09167 & Pass\\
$X_5\left(10\right)$ & $\left | X_5\left(\cdot\right) \right | < 2.326348$ & 1.288804 & Pass & 0 & Pass\\
$X_5\left(11\right)$ & $\left | X_5\left(\cdot\right) \right | < 2.326348$ & 0.09245 & Pass & -0.64715 & Pass\\
$X_5\left(12\right)$ & $\left | X_5\left(\cdot\right) \right | < 2.326348$ & 0 & Pass & 0 & Pass\\
$X_5\left(13\right)$ & $\left | X_5\left(\cdot\right) \right | < 2.326348$ & 0.466252 & Pass & -0.466252 & Pass\\
$X_5\left(14\right)$ & $\left | X_5\left(\cdot\right) \right | < 2.326348$ & 0.749269 & Pass & -0.561951 & Pass\\
$X_5\left(15\right)$ & $\left | X_5\left(\cdot\right) \right | < 2.326348$ & -0.658505 & Pass & -0.282216 & Pass\\
$X_5\left(16\right)$ & $\left | X_5\left(\cdot\right) \right | < 2.326348$ & 0.188982 & Pass & -0.377964 & Pass\\
$X_5\left(17\right)$ & $\left | X_5\left(\cdot\right) \right | < 2.326348$ & 1.233905 & Pass & -1.993232 & Pass\\
$X_5\left(18\right)$ & $\left | X_5\left(\cdot\right) \right | < 2.326348$ & 0.381385 & Pass & -2.097618 & Pass\\
$X_5\left(19\right)$ & $\left | X_5\left(\cdot\right) \right | < 2.326348$ & -0.287348 & Pass & -0.478913 & Pass\\
$X_5\left(20\right)$ & $\left | X_5\left(\cdot\right) \right | < 2.326348$ & -1.539601 & Pass & -0.96225 & Pass\\
$X_5\left(21\right)$ & $\left | X_5\left(\cdot\right) \right | < 2.326348$ & 1.06341 & Pass & -0.290021 & Pass\\
$X_5\left(22\right)$ & $\left | X_5\left(\cdot\right) \right | < 2.326348$ & 0.194257 & Pass & -1.748315 & Pass\\
$X_5\left(23\right)$ & $\left | X_5\left(\cdot\right) \right | < 2.326348$ & 1.85421 & Pass & -2.04939 & Pass\\
$X_5\left(24\right)$ & $\left | X_5\left(\cdot\right) \right | < 2.326348$ & 0.784465 & Pass & 0.392232 & Pass\\
$X_5\left(25\right)$ & $\left | X_5\left(\cdot\right) \right | < 2.326348$ & 1.280928 & Pass & 0.68973 & Pass\\
$X_5\left(26\right)$ & $\left | X_5\left(\cdot\right) \right | < 2.326348$ & 1.782266 & Pass & -0.594089 & Pass\\
$X_5\left(27\right)$ & $\left | X_5\left(\cdot\right) \right | < 2.326348$ & -0.298511 & Pass & -0.298511 & Pass\\
$X_5\left(28\right)$ & $\left | X_5\left(\cdot\right) \right | < 2.326348$ & -0.4 & Pass & -1.8 & Pass\\
$X_5\left(29\right)$ & $\left | X_5\left(\cdot\right) \right | < 2.326348$ & 1.306549 & Pass & 1.105542 & Pass\\
$X_5\left(30\right)$ & $\left | X_5\left(\cdot\right) \right | < 2.326348$ & 0.808122 & Pass & 0.606092 & Pass\\
$X_5\left(31\right)$ & $\left | X_5\left(\cdot\right) \right | < 2.326348$ & 1.929158 & Pass & -0.507673 & Pass\\
$X_5\left(32\right)$ & $\left | X_5\left(\cdot\right) \right | < 2.326348$ & 0 & Pass & -0.408248 & Pass\\
$X_5\left(33\right)$ & $\left | X_5\left(\cdot\right) \right | < 2.326348$ & -0.102598 & Pass & 0.307794 & Pass\\
$X_5\left(34\right)$ & $\left | X_5\left(\cdot\right) \right | < 2.326348$ & 0 & Pass & -0.206284 & Pass\\
$X_5\left(35\right)$ & $\left | X_5\left(\cdot\right) \right | < 2.326348$ & 0.933257 & Pass & -2.384989 & Fail\\
$X_5\left(36\right)$ & $\left | X_5\left(\cdot\right) \right | < 2.326348$ & 0.625543 & Pass & -0.208514 & Pass\\
$X_5\left(37\right)$ & $\left | X_5\left(\cdot\right) \right | < 2.326348$ & -0.314485 & Pass & -0.943456 & Pass\\
$X_5\left(38\right)$ & $\left | X_5\left(\cdot\right) \right | < 2.326348$ & -0.843274 & Pass & -1.897367 & Pass\\
$X_5\left(39\right)$ & $\left | X_5\left(\cdot\right) \right | < 2.326348$ & -0.317999 & Pass & -0.317999 & Pass\\
$X_5\left(40\right)$ & $\left | X_5\left(\cdot\right) \right | < 2.326348$ & -1.918806 & Pass & -0.639602 & Pass\\
$X_5\left(41\right)$ & $\left | X_5\left(\cdot\right) \right | < 2.326348$ & 1.393746 & Pass & -0.964901 & Pass\\
$X_5\left(42\right)$ & $\left | X_5\left(\cdot\right) \right | < 2.326348$ & -0.862662 & Pass & 0.215666 & Pass\\
$X_5\left(43\right)$ & $\left | X_5\left(\cdot\right) \right | < 2.326348$ & -0.325396 & Pass & -0.325396 & Pass\\
$X_5\left(44\right)$ & $\left | X_5\left(\cdot\right) \right | < 2.326348$ & -1.309307 & Pass & -0.872872 & Pass\\
$X_5\left(45\right)$ & $\left | X_5\left(\cdot\right) \right | < 2.326348$ & 1.426935 & Pass & -1.426935 & Pass\\ 
$X_5\left(46\right)$ & $\left | X_5\left(\cdot\right) \right | < 2.326348$ & -0.441726 & Pass & -0.220863 & Pass\\
$X_5\left(47\right)$ & $\left | X_5\left(\cdot\right) \right | < 2.326348$ & -1 & Pass & 0.777778 & Pass\\
$X_5\left(48\right)$ & $\left | X_5\left(\cdot\right) \right | < 2.326348$ & -0.67082 & Pass & -0.447214 & Pass\\
$X_5\left(49\right)$ & $\left | X_5\left(\cdot\right) \right | < 2.326348$ & -1.237597 & Pass & -1.012579 & Pass\\
$X_5\left(50\right)$ & $\left | X_5\left(\cdot\right) \right | < 2.326348$ & 0.452911 & Pass & 0.679366 & Pass\\
$X_5\left(51\right)$ & $\left | X_5\left(\cdot\right) \right | < 2.326348$ & -1.025645 & Pass & 0.797724 & Pass\\
$X_5\left(52\right)$ & $\left | X_5\left(\cdot\right) \right | < 2.326348$ & -0.458831 & Pass & 0.458831 & Pass\\
$X_5\left(53\right)$ & $\left | X_5\left(\cdot\right) \right | < 2.326348$ & -0.11547 & Pass & -0.57735 & Pass\\
$X_5\left(54\right)$ & $\left | X_5\left(\cdot\right) \right | < 2.326348$ & -0.464991 & Pass & -1.394972 & Pass\\
$X_5\left(55\right)$ & $\left | X_5\left(\cdot\right) \right | < 2.326348$ & -0.819288 & Pass & -0.117041 & Pass\\
$X_5\left(56\right)$ & $\left | X_5\left(\cdot\right) \right | < 2.326348$ & 0.235702 & Pass & -0.471405 & Pass\\
$X_5\left(57\right)$ & $\left | X_5\left(\cdot\right) \right | < 2.326348$ & -1.780172 & Pass & 0.593391 & Pass\\
$X_5\left(58\right)$ & $\left | X_5\left(\cdot\right) \right | < 2.326348$ & -0.717137 & Pass & 0.239046 & Pass\\
$X_5\left(59\right)$ & $\left | X_5\left(\cdot\right) \right | < 2.326348$ & -1.324244 & Pass & -1.083473 & Pass\\
$X_5\left(60\right)$ & $\left | X_5\left(\cdot\right) \right | < 2.326348$ & -0.242536 & Pass & 0.242536 & Pass\\
$X_5\left(61\right)$ & $\left | X_5\left(\cdot\right) \right | < 2.326348$ & 1.588203 & Pass & -0.855186 & Pass\\
$X_5\left(62\right)$ & $\left | X_5\left(\cdot\right) \right | < 2.326348$ & -0.246183 & Pass & -1.723281 & Pass\\
$X_5\left(63\right)$ & $\left | X_5\left(\cdot\right) \right | < 2.326348$ & 0.620174 & Pass & 0.124035 & Pass\\
$X_5\left(64\right)$ & $\left | X_5\left(\cdot\right) \right | < 2.326348$ & -0.5 & Pass & 0.5 & Pass\\

 \hline
		\end{tabular} \end{scriptsize}
	\end{center}
	\caption{Testing for Menezes' Basic Tests of Randomness}
	\label{tab:Menezes}
\end{table}

\subsection{FIPS 140-1}
Testing for compliance with FIPS 140-1 is required to be carried out with sequences of 20,000 bits.  Five sequences $S_1$ through $S_5$ were generated using parameters shown in table \ref{tab:FIPSParameters}.  The results of the tests on the sequences are shown in table \ref{tab:FIPS-140-1}.
\begin{table}[ht]
		\begin{center} 
		\begin{tabular}{|l|l|l|l|l|l|}
		\hline
			\textbf{Parameter}	&					$\mathbf{S_1}$ & $\mathbf{S_2}$ & 	$\mathbf{S_3}$ & $\mathbf{S_4}$ & 	$\mathbf{S_5}$ \\ \hline
			$\alpha$ &	1.23 &	1.40 &	1.40 &	1.40 &	1.41 \\
			$\beta$ &		0.25 &	0.25 &	0.30 &	0.30 &	0.21 \\
			$x_0$ &			-1.0 &	-1.0 &	-1.0 &	-1.0 &	-1.0 \\
			$y_0$ &			1.0 &		1.0 &		1.0 &		1.0 &		1.0 \\
			$B_y\left(-2\right)$ & 0 & 0 & 0 & 0 & 0 \\
			$B_y\left(-1\right)$ & 1 & 1 & 1 & 1 & 1 \\
			$P$ & 84 & 84 & 84 & 24 & 24 \\
			$T$ & 1000 & 1000 & 1000 & 1000 & 1000\\ \hline
		\end{tabular}
	\end{center}
	\caption{Parameters used for FIPS 140-1 test samples}
	\label{tab:FIPSParameters}
\end{table}
\begin{table}[ht]
		\begin{center} 
		\begin{tabular}{|l|l|l|l|l|l|l|}
		\hline
			\textbf{Statistic}	& \textbf{Expected} &					$\mathbf{S_1}$ & $\mathbf{S_2}$ & 	$\mathbf{S_3}$ & $\mathbf{S_4}$ & 	$\mathbf{S_5}$ \\ \hline
			$n_1$						&	$9654<n_1<10346$ & 	9938 &		10107 & 		9944 &		10099 & 		10020 \\
			$X_3$						&	$1.03<X_3<57.40$ & 	17.25 &	13.03 & 	12.98 &	14.73 & 	6.66\\
			$B_1$						&	$2267<B_1<2733$ & 	2572 &		2473 & 			2560 &		2480 & 			2454\\ 
			$G_1$						&	$2267<G_1<2733$ & 	2452 &		2524 & 			2534 &		2554 & 			2447\\ 
			$B_2$						&	$1079<B_2<1421$ & 	1192 &		1231 & 			1200 &		1286 & 			1268\\
			$G_2$						&	$1079<G_2<1421$ & 	1302 &		1264 & 			1226 &		1262 & 			1310\\
			$B_3$						&	$502<B_3<748$ & 		640 &			643 & 			653 &			636 & 			616\\
			$G_3$						&	$502<G_3<748$ & 		611 &			577 & 			638 &			582 & 			580\\
			$B_4$						&	$223<B_4<402$ & 		277 &			319 & 			320 &			312 & 			312\\
			$G_4$						&	$223<G_4<402$ & 		328 &			315 & 			306 &			336 & 			314\\
			$B_5$						&	$90<B_5<223$ & 			140 &			165 & 			162 &			150 & 			156\\
			$G_5$						&	$90<G_5<223$ & 			156 &			165 & 			168 &			148 & 			159\\
			$B_6$						&	$90<B_6<223$ & 			180 &			159 & 			134 &			159 & 			164\\
			$G_6$						&	$90<G_6<223$ & 			151 &			146 & 			157 &			142 & 			159\\ \hline
			Result					& Pass/Fail &					Pass &		Pass & 			Pass &		Pass &			Pass\\ \hline
		\end{tabular}
	\end{center}
	\caption{Testing for FIPS 140-1}
	\label{tab:FIPS-140-1}
\end{table}

\subsection{NIST Statistical Test Suite}
For each test-run of the test-suite, a sequence of $2 \times 10^{8}$ bits was generated and the test-suite was configured to consider this sequence as 200 sequences of $1 \times 10^{6}$ bits each.  This set of $200$ sequences was subjected to statistical testing in each case.  In this manner, the distribution of failures could also be examined by the test-suite and appropriate analysis could be carried out.

The test was carried out on two sets $U_1$ and $U_2$ of 200-samples each.  The parameters used for generating these sets of sequences are shown in table \ref{tab:NISTParameters}.
\begin{table}[ht]
		\begin{center} 
		\begin{tabular}{|l|l|l|}
		\hline
			\textbf{Parameter}	&					$\mathbf{U_1}$ & $\mathbf{U_2}$ \\ \hline
			$\alpha$ &	1.40 &	1.398 \\
			$\beta$ &		0.30 &	0.283 \\
			$x_0$ &			-1 &	0.26 \\
			$y_0$ &			1 &		0.29 \\
			$B_y\left(-2\right)$ & 0 & 0 \\
			$B_y\left(-1\right)$ & 1 & 1 \\
			$P$ & 117 & 111 \\
			$T$ & 1000 & 1000 \\ \hline
		\end{tabular}
	\end{center}
	\caption{Parameters used for NIST test samples}
	\label{tab:NISTParameters}
\end{table}

For a sample of 200 sequences, the minimum proportion of sequences required to pass all the tests of the suite other than the random excursions (variant) test is 0.968893.  The proportion of sequences passing these tests was found to be larger than this threshold.  For the random excursions (variant) test, the required minimum passing proportion was found to be 0.961540 and 0.962864 for $U_1$ and $U_2$ respectively.  $U_1$ and $U_2$ were found to meet this requirement also.  The author has noticed, however, that some of the sequences that pass the FIPS 140-1 and Menezes' tests do not pass the NIST suite of tests.  For example, a set of sequences generated using the same parameters as $R_1$ was found to fail in the NIST suite.  The author has found that passing the NIST suite requires a more careful choice of the parameters.  In this case, increasing the value of $T$ to a sufficiently large value resulted in the sequences passing the NIST suite.

\section{Keyspace Size}
The properties of H\'enon map sequences presented in the preceding sections demonstrate that they are potential candidates for cryptographic applications.  A H\'enon map sequence can be used, for example, can be directly Exclusive-ORed, bit-by-bit, with a data sequence of the same length.  Such a cipher is popularly known as the Vernam Cipher \cite{kippenhahn99,mollin01}.  The values of $\alpha, \beta, x_0, y_0$ and the sampling-factor $P$ together can form the key.  In this section, an attempt is made to estimate the size of the keyspace for such a cipher.

The author has found that $\alpha$ in $\left(1.16, 1.41\right)$ and $\beta$ in $\left(0.2, 0.3\right)$ are useful for generating sequences with the desired statistical properties.  Also, we can assume $x_0$ to lie in $\left(-1,1\right)$ and $y_0$ to lie in $\left(-0.35, 0.35\right)$.  $P$ can be assumed within $\left(80,1000\right)$.  Though these limits are in no way binding or accurate, they should give a reasonable estimate of the size of the keyspace.  The size of the keyspace then depends on the precision of the computing platform on which the cipher algorithm is implemented.  With 32-bit floating-point numbers of the IEEE format, the smallest possible increment $\left(\epsilon\right)$ is $\epsilon_{32} \approx 1.1921 \times 10^{-7}$.  With 64-bit floating-point numbers this value is $\epsilon_{64} \approx 2.2204 \times 10^{-16}$.  Let $\hat{\alpha}$ indicate the size of the interval over which $\alpha$ can span i.e. $\hat{\alpha} = 1.41 - 1.16$.  Let $\hat{\beta}, \hat{x_0}, \hat{y_0}$ and $\hat{P}$ have correspondingly similar meaning.  Since $P$ is an integer, $\hat{P} = 1000 - 80$.  Then, the number of representable values of $\alpha$ on the applicable computing platform can be computed as $K_{\alpha} = \hat{\alpha}/\epsilon$.  Such a calculation can easily be carried out using logarithms.  Similarly, $K_{\beta}, K_{x_0}$ and $K_{y_0}$ can also be calculated.  $K_P = \hat{P}$.  Since the parameters are used together, the size of the keyspace $K = K_{\alpha} \times K_{\beta} \times K_{x_0} \times K_{y_0} \times K_P$.  The logarithm of this figure, to the base 2, gives an estimate of the length of a single binary key which contains all the information required to generate the pseudo-random sequence.  The value of $\log_2\left(K\right)$ for 32-bit and 64-bit precision was found to be 97 and 213 respectively.  The size of the keyspace can be further increased by increasing the precision of floating-point representation.

\section{Concluding Remarks}
Though chaotic orbits of discrete-time maps are non-periodic in nature, because of finite precision of digital computers the orbits actually turn out to be periodic.  The average period of an orbit of a two-dimensional map can be expected to be longer than that of a one-dimensional map.  To overcome the inevitable periodicity in digital computation, Shujun et al have used simple LFSR-based perturbation generators to perturb the parameters of the dynamical systems used \cite{shujun01}.  Though they have used perturbed one-dimensional maps in couple-chaotic-system-based pseudorandom sequence generators, the same technique should be directly applicable to the present algorithm and is a feasible way of defeating the periodicity.

The periodicity inherent in digital-computer implementations is not a problem in maps realized on an analog-computer.  The H\'enon map, due to its polynomial form, can be realized in the form of an electronic circuit using analog-multipliers, sample-and-hold blocks and operational amplifiers.  Such a realization of the logistic map has already been studied \cite{suneel06}.  The advantage of an analog realization of chaotic maps is that due to natural variations in circuit parameters and conditions, practically identical systems provided with practically identical conditions generate sequences that quickly diverge and become un-correlated within a short time.  Therefore, such implementations may actually turn out to be \emph{truly} random (as opposed to pseudorandom) sequence generators.

The choice of the H\'enon map for the work in this paper was rather arbitrary.  The author believes that similar results should also be attainable with other two-dimensional maps.

The linear complexity and correlation properties of the proposed sequences suggest a strong similarity to random sequences.  Results of statistical tests carried out confirm this further.  A large size of the keyspace suggests strong candidature for cryptographic applications, especially for stream cipher cryptography.  Possible cryptanalysis techniques for these sequences is an open subject.

\section{Acknowledgements}
The author thanks his colleagues A.P. Dabhade, K.V. Suresh, D. Venu Gopal and R.S. Chandrasekhar for several hours of  discussions.

%
%

\end{document}